\documentclass[a4paper]{jpconf}
\usepackage{blindtext}
\usepackage{subfigure}

\usepackage{graphicx}
\begin{document}
\title{Identifying the nature of high energy Astroparticles}

\author{Karen Salom\'e Caballero Mora}

\address{Full time Professor, Facultad de Ciencias en F\'isica y Matem\'aticas, Universidad Aut\'onoma de Chiapas, Carretera Emiliano Zapata Km. 8 Rancho San Francisco, Ciudad
Universitaria, Ter\'an, Tuxtla Guti\'errez, Chiapas, Mexico}

\ead{karen.mora@unach.mx}

\begin{abstract}
\noindent High energy Astroparticles include Cosmic Ray (CR), gamma ray and neutrinos, all of them coming from the universe. The origin and production, acceleration and propagation mechanisms of ultrahigh-energy CR (UHECR $\sim 10^{20}$ eV) are still unknown. Knowledge on particle interactions taking place at those energies, useful for studying current theories on particle physics, can be obtained only from measurements of high energy astroparticles. In the present document some techniques on data analysis of mass composition of UHECR with the Pierre Auger Observatory are described. The relevance of the muon component of air showers produced by the primary CR, as well as some low energy simulations of that component, are explained. 
\end{abstract}
\vspace{-1 cm}
\section{Introduction}

\begin{figure}[h]
\hspace{-1.0pc}
\begin{minipage}{22.0pc}
\includegraphics[width=22.0pc]{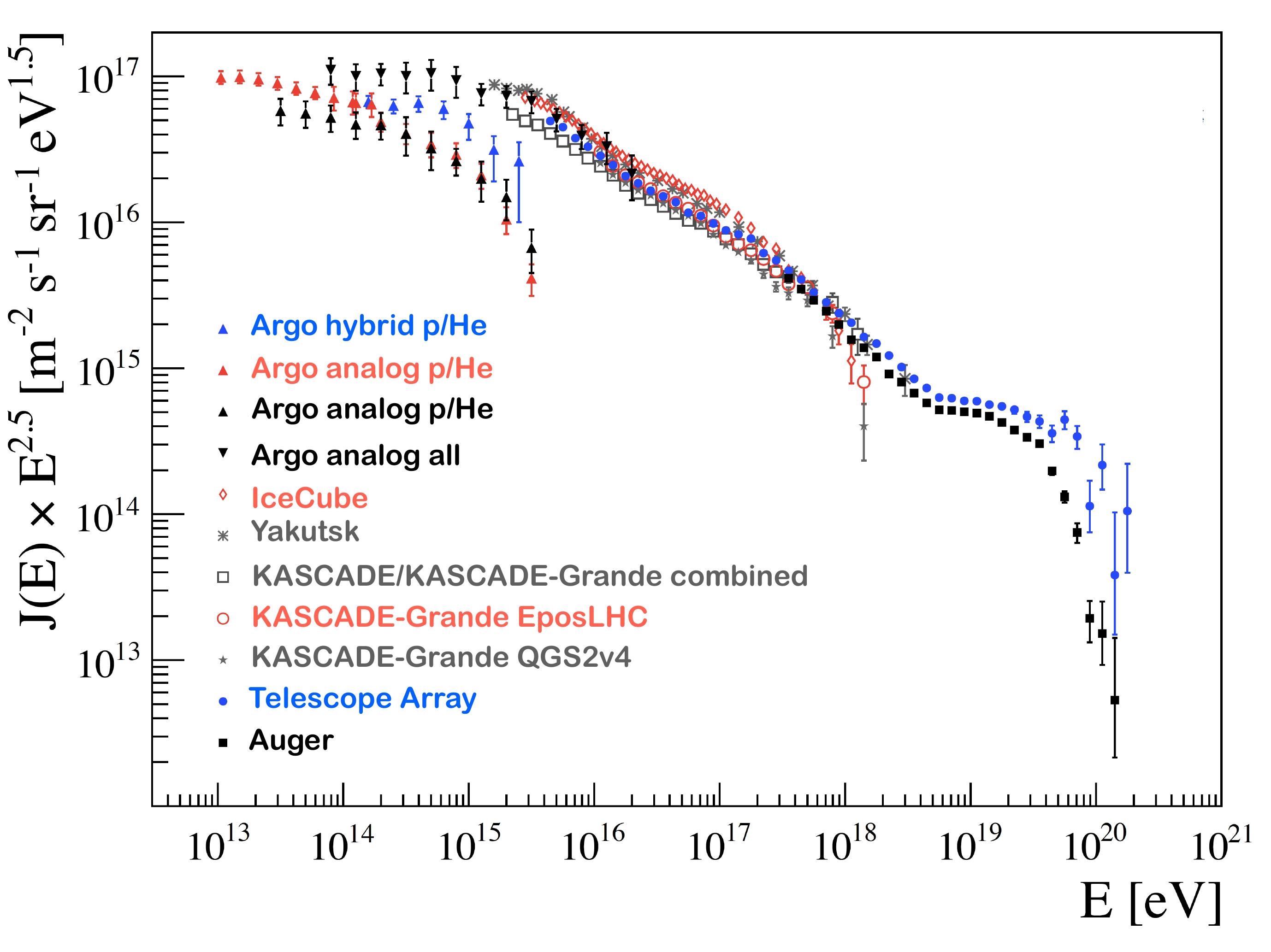}
\begin{center}
\vspace{-0.5 cm}
\caption{\label{fig:Spectrum}Cosmic ray energy spectrum as measured by many experiments over a wide range in energy, as presented at ICRC 2015 (modified from \cite{ValerioICRC}).}
\end{center}
\end{minipage}\hspace{0.5pc}%
\begin{minipage}{17pc}
\includegraphics[width=17pc]{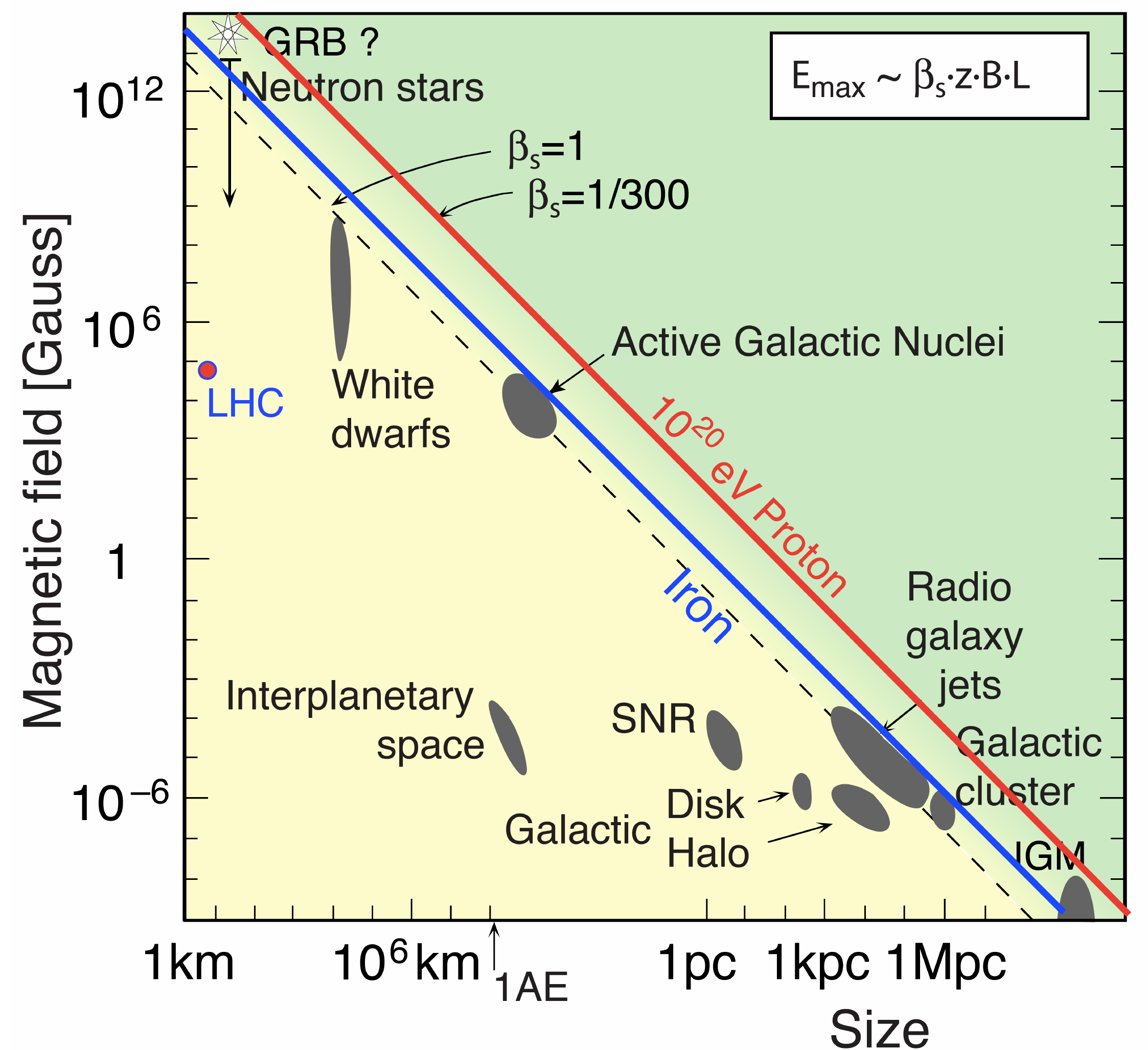}
\begin{center}
\vspace{-0.5 cm}
\caption{\label{fig:Hillasplot}Hillas Plot showing the required size and magnetic field strength of possible UHECR sources. Objects below the corresponding lines are not able to accelerate proton or iron nuclei to an energy of $10^{20}$eV \cite{REngel}.}
\end{center}
\end{minipage} 
\vspace{-0.8 cm}
\end{figure}

\noindent The energy spectrum of CR shows the flux of particles as a function of energy. It extends from a few GeV to above $10^{20}$ eV. Acceleration mechanism together with propagation processes through the interstellar medium can explain qualitatively the whole energy range but the acceleration sites are still to be explained. The energy spectrum has some features which are easily distinguishable because of its similarity with a human leg, the knee and the ankle. Those features might be interpreted either as a change of the acceleration mechanism at sources, as a propagation effect or as a change of the hadronic interaction cross sections with increasing energy \cite{KarenPhD}. Figure~\ref{fig:Spectrum} shows the spectrum measured by different experiments \cite{ValerioICRC}. 
\noindent CR with energies up to some GeV arise predominantly from the Sun. From these energies up to a few $10^{15}$ eV CR could be originated by shock acceleration in galactic  Supernova Remnants (SNRs)~\cite{Aharonian_2004vr}. Although there is no experimental evidence, some theories point out that  SNRs could accelerate cosmic rays up to energies of $10^{18}$ eV  as well~\cite{Ptuskin,Hillas_2005cs}. The maximum energy that CR may obtain from a given astronomical object is limited by the combination of the magnetic field strength and shock size \cite{Drury_1994fg,Nagano_2000ve}. On this basis, M. Hillas proposed a simple plot (Figure~\ref{fig:Hillasplot}) to show that there are only few astrophysical objects (Active Galactic Neclei-AGNs and Gamma Ray Burst-GRBs, Intergalactic Medium-IGM is also shown) capable to accelerate nuclei and protons to energies of $10^{20}$ eV. There are other non-traditional theories trying to explain the origin of the extreme energetic particles, the so called {\it top-down} models. Some of those models include production of UHECRs from the decay of exotic particles such as Topological Defects (TD)~\cite {Hill_1982iq}, super heavy dark matter (SHDM) particles ~\cite{Berezinsky_1997hy,Aloisio_2003xj,Ellis_2005jc}, or neutrino interactions with the relic neutrino background (Z-burst model)~\cite {weiler-1999-11,Fargion_1997ft}. All the top-down models suggest a large fraction of photons in the flux of UHECRs~\cite {Risse_2007sd}, which can be studied measuring the flux limits for photons and neutrinos. The Pierre Auger Collaboration has estimated upper limits on both of them~\cite {Auger_gamma,Auger_neutrino}. The upper limit set to the flux of photons allowed to constrain certain top-down models, which can not be considered as plausible scenarios for the production of UHECRs any more. The GZK effect \cite{gzk1,gzk2} also imposes a limit for the distance at which the sources of cosmic rays above energies of $5\cdot10^{19}$ eV can be located. Identification of sources of UHECRs by looking directly at their arrival directions is not easy because at relative low energies, the distribution of the arrival directions is isotropic. Nevertheless, a lot of effort has been concentrated to find significant anisotropies.
\noindent Most of high energy particles in CR are protons; about $10\%$ are helium nuclei, and $1\%$ are neutrons or nuclei of heavier elements. Together, these account for $99\%$ of the CR, and electrons and photons conform the remaining $1\%$. The number of neutrinos is estimated to be comparable to that of high energy photons, but it is very high at low energy because of the nuclear processes that occur in the Sun \cite{MPimenta}. For UHECRs, studies on the depth of shower maximum performed by the Pierre Auger Observatory, have recently shown an unexpected non-trivial behavior in the mass composition above $10^{18}$ eV, more information on these results is described in this document. Interpreted with the leading LHC-tuned shower models, this behavior implies a gradual shift to a heavier composition. A number of fundamentally different astrophysical model scenarios have been developed to describe such evolution \cite{AugerPrime}.

\noindent When CR arrive to the atmosphere, they produce an extensive air shower (EAS) of secondary particles. They reach a maximum and continues after the total energy of the primary cosmic ray is distributed to them ~\cite{gaisser}. The EAS has the components: {\it hadronic}, {\it muonic} and {\it electromagnetic}. In addition, there are particles with small contribution to the total energy balance, i.e. {\it UV-photons} and {\it radio emission}, or which are not detectable called {\it invisible component} (neutrinos and very low energy particles)~\cite{RalfUlrichPhD}. Precision Auger measurements of shower properties, have revealed inconsistencies within present shower models, opening the possibility that the unexpected behavior is due to new hadronic interaction physics at energy scales beyond the reach of the LHC \cite{AugerPrime}. Gamma photons are photons of very high energy which can travel long distances without being deflected by galactic and extragalactic magnetic fields, hence they allow us to directly study their emission sources \cite{MPimenta}, neutrinos have the same situation. Gamma rays are involved in phenomena taking place in astrophysical objects such as AGNs and GRBs, as well as Dark Matter annihilation and decay, Lorentz Invariance Violation (LIV) and inter-galactic magnetic field measurement \cite{Rlauer}. Such emissions can be meassured with HAWC experiment at TeV-scale \cite{ASmithSISSA}. There is a feedback between CR physics and particle physics trying to understand the mechanisms of their interactions with matter and giving the possibility to develop fundamental theories of physics \cite{24AugerPrime, 25AugerPrime, 26AugerPrime, AugerPrime}. UHERC are an alternative field of research to look for new physics beyond the Standard Model of elementary particles, in a kinematic region not accesible by terrestrial accelerators on the basic idea of the modification of effective cross sections at high energies. Examples of such theories are Supersymmetry (SUSY), Grand Unified Theories (GUT's) \cite{2FisNva} and Extra Dimensions Models \cite{3FisNva}.	
\vspace{-0.5 cm}
\section{Mass Composition with the Pierre Auger Observatory}
The Pierre Auger Observatory, is an international collaboration conformed by $\sim$500 people from 82 institutions in 16 countries \cite{AugerWeb}. The scientific goal of Auger is to study UHECR ($10^{18} $ to $10^{20}$ eV). The most important achievements of the experiment are described in \cite{AugerPrime}
\vspace{-0.5 cm}
\subsection{Description and performance of the Observatory}
\label{PerformanceAuger}
Auger is a hybrid detector, it covers an area of 3000 $km^{2}$. The original components are a surface detector (SD) with a duty cycle of $100\%$ and a fluorescence detector (FD) with a duty cycle of $15\%$.  The \textbf{SD} has 1660 3.6 m diameter water Cherenkov tanks. Each station is full with 12,000 $l$ of pure water. The light produced inside the tank is measured by three 9''Photonis photomultipliers tubes (PMTs). The PMTs are set on top of the tank symmetrically distributed, they look downwards into the water. The SD observes the particles of the EAS arriving to the earth, recording the lateral distribution. There are four \textbf{FD} stations located on the edges of the SD grid. Each of the stations contain six individual air-fluorescence light telescopes consisting of a spherical mirror with a spherical pixel camera. The camera is equiped with 440 hexagonal Photonis PMTs. The FD observes the longitudinal development of the EAS at the atmosphere.
\noindent \textbf{Extensions} With the goal of studying the transition of galactic to extragalactic CR and to better understand the composition, Auger has included some enhancements including high elevetion fluorescence telescopes (HEAT), the AMIGA underground muon detectors and detectors of radio (AERA) and microwave emissions from air showers \cite{AugerPrime}.
\noindent \textbf{Auger Prime (Primary cosmic Ray Identification with Muons and Electrons)}
\noindent It consists of an upgrade of the experiment. Motivations: to get information on mass composition and the origin of the flux suppression at the highest energies, the search of a flux contribution of protons up to  the highest energies, to study EAS and hadronic multiparticle production. The upgrades consist of a complementary  Surface Scintilator Detector (SSD) above the existing SD stations, substitution of current electronics, the underground muon detector AMIGA  and the extension of  duty cycle of FD \cite{AugerPrime}.
\vspace{-0.5 cm}
\subsection{Mass Composition sensitive parameters}
\label{Mass}
To know the mass composition of UHECR could give valuable information on the origin and acceleration mechanisms of such particles, as well as set constrains to the models. Interpreting the observed longitudinal shower profiles \cite{22AugerPrime} with LHC-tuned interaction models, the Pierre Auger Observatory can conclude that there is a large fraction of protons present at $10^{18}$ eV, changing to a heavier composition, possibly dominated by elements of the CNO group, at $10^{19.5}$ eV \cite{22AugerPrime}. In this section one of the first results (with data up to 2009) supporting this finding, based on mean values of $X_{max}$ and Risetime mass composition sensitive parameters is described. \textbf{Risetime} ( {\boldmath $t_{1/2}$}) is the time it takes for the integrated signal recorded by the SD stations, to rise from $10\%$ to $50\%$ of the final value. It is sensitive to the mass composition since its value changes for light initiated EAS against that of heavy initiated ones \cite{Nagano_2000ve,Barnhill}. Parameters derived from $t_{1/2}$ as its asymmetry \cite{69AugerPrime},  $\Delta_{i}$ \cite{BSmith} and $\Delta_{1000}$ are also mass sensitive. {\boldmath $X_{max}$} is the most mass-sensitive observable recorded by the FD, it is the depth of the shower maximum, meassured by looking at the longitudinal shower profile \cite{22AugerPrime}.
\vspace{-0.5 cm}
\subsection{Method for estimation of mass composition at the highest available energies} 
\label{Risetimeresults}
Observable $\Delta_{1000}$ is calculated for high quality hybrid events and calibrated with $X_{max}$. With this method one takes advantage of the high statistics provided by the SD to obtain a new estimation for $X_{max}$, $X^{SD} _{max}$. The energy range of the Observatory for measuring $X_{max}$ was extended to energies of 80 EeV. The mass composition corresponding to energies above $10^{19.6}$ eV is found to get heavier but still remains in a mixed mass composition, if the evolution of mean $X_{max}$ with the energy is compared with models \cite{KarenPhD}.
\begin{figure}
\hspace{-0.8pc}
  \begin{minipage}[b]{0.99\linewidth}
\begin{tabular}{llll}
\br
Zenith Angle &  Energy &InD\\
($\circ$)&x $10^{19}$ (eV)&(m)\\
\mr
36.56&2.29&765\\
44.77&1.37&910\\
57.57&1.62&1095\\
\br
\end{tabular}
\centering
\caption{\label{ex1}  \vspace{3pt}  Examples of characteristic distances for $t_{1/2}$ for different zenith angles}
\par\vspace{0pt}
\end{minipage}
\end{figure}

\noindent \textbf{Improvement of $t_{1/2}$ parameter} Typically $t_{1/2}$ is considered at 1000 m from the shower core per event. Nevertheless this parameter presents a large spread \cite {KarenPhD}. Since there is no physical reason to take that distance, it is worth to explore if there is other characteristic distance for each different event. Currently a study to find such a distance is being performed. It is found as the intersection distance (InD) of several parameterisations of $t_{1/2}$ as a function of the distance to the shower core for a given event. This is similar to the method used to find the 1000 m distance for the Lateral Distribution Function \cite{Ioana} .  It is found, after analysing 15 high quality events, that the interesection changes for different zenith angles, preliminary representative results are shown in Figure \ref{ex1}. More studies are in progress \cite{Hernan}.
\vspace{-0.5 cm}
\subsection{Muon Component}
\label{MuonComp}
The muonic component of air showers is sensitive to hadronic particle interactions at all stages in the EAS, and to many properties of hadronic interactions such as the multiplicity, elasticity, fraction of neutral secondary pions, and the baryon-to-pion ratio. Currently the number of muons can only be measured indirectly except at very large lateral distances and in very inclined showers, where muons dominate the shower signal at ground level \cite {AugerPrime}. The muon content in EAS also gives information on the composition of the primary UHECR, an iron primary may induce up to about $40 \%$ more muons than a proton primary of the same energy \cite{MiProceeding}. 

\begin{figure}[h]
\hspace{-0.8pc}
\begin{minipage}{20pc}
\includegraphics[width=20pc]{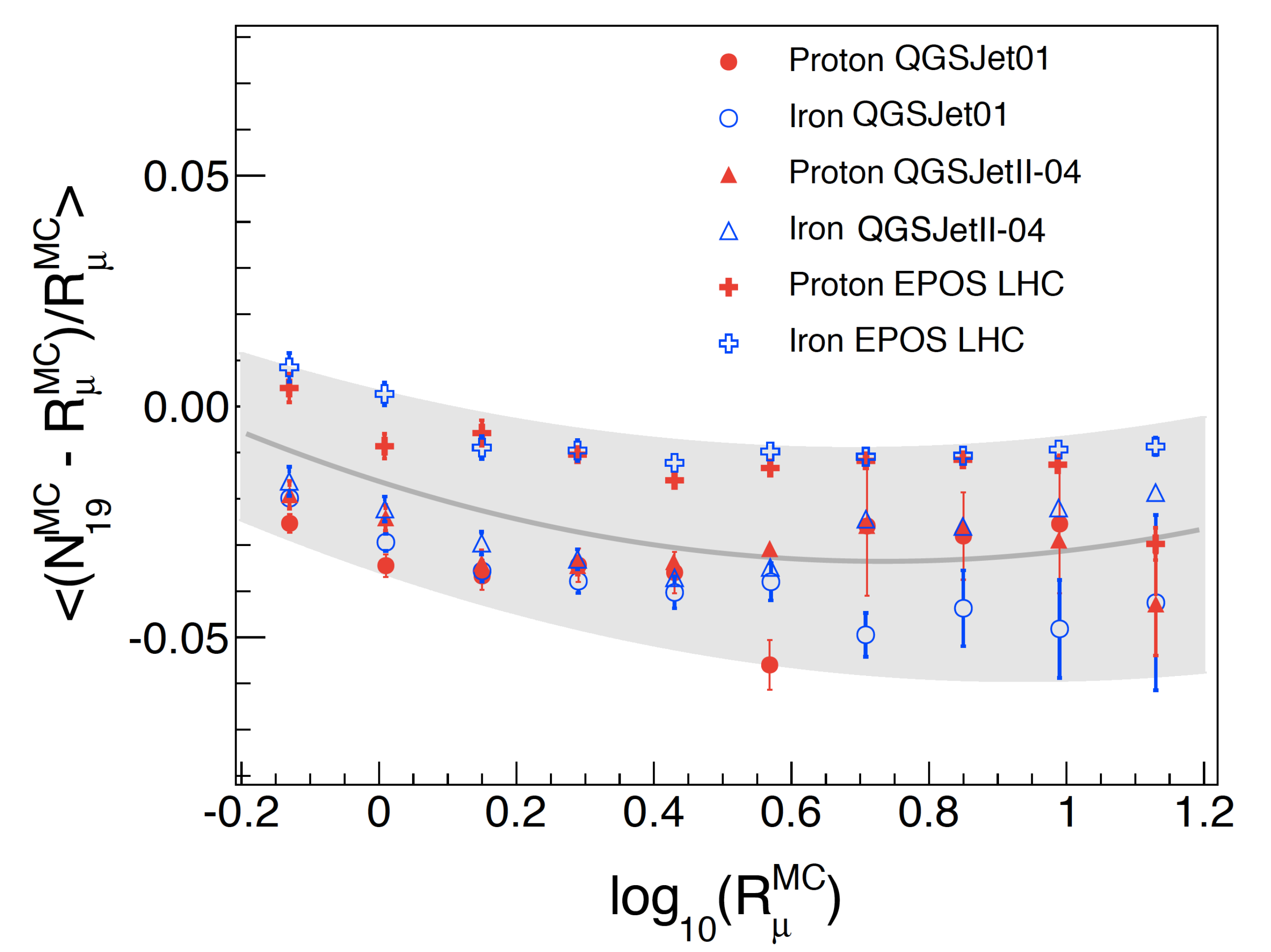}
\begin{center}
\vspace{-0.5 cm}
\caption{\label{fig:MuonInclined}Average of the relative difference between $N^{MC} _{19}$  and $R^{MC} _{\mu}$ for proton and iron showers simulated with QGSJet01, QGSJetII-04 and EPOS LHC. The band indicates the bias region and the solid line indicates the parameterised average bias \cite{MiInclinada}.}
\end{center}
\end{minipage}\hspace{0.8pc}%
\begin{minipage}{16pc}
\includegraphics[width=17pc]{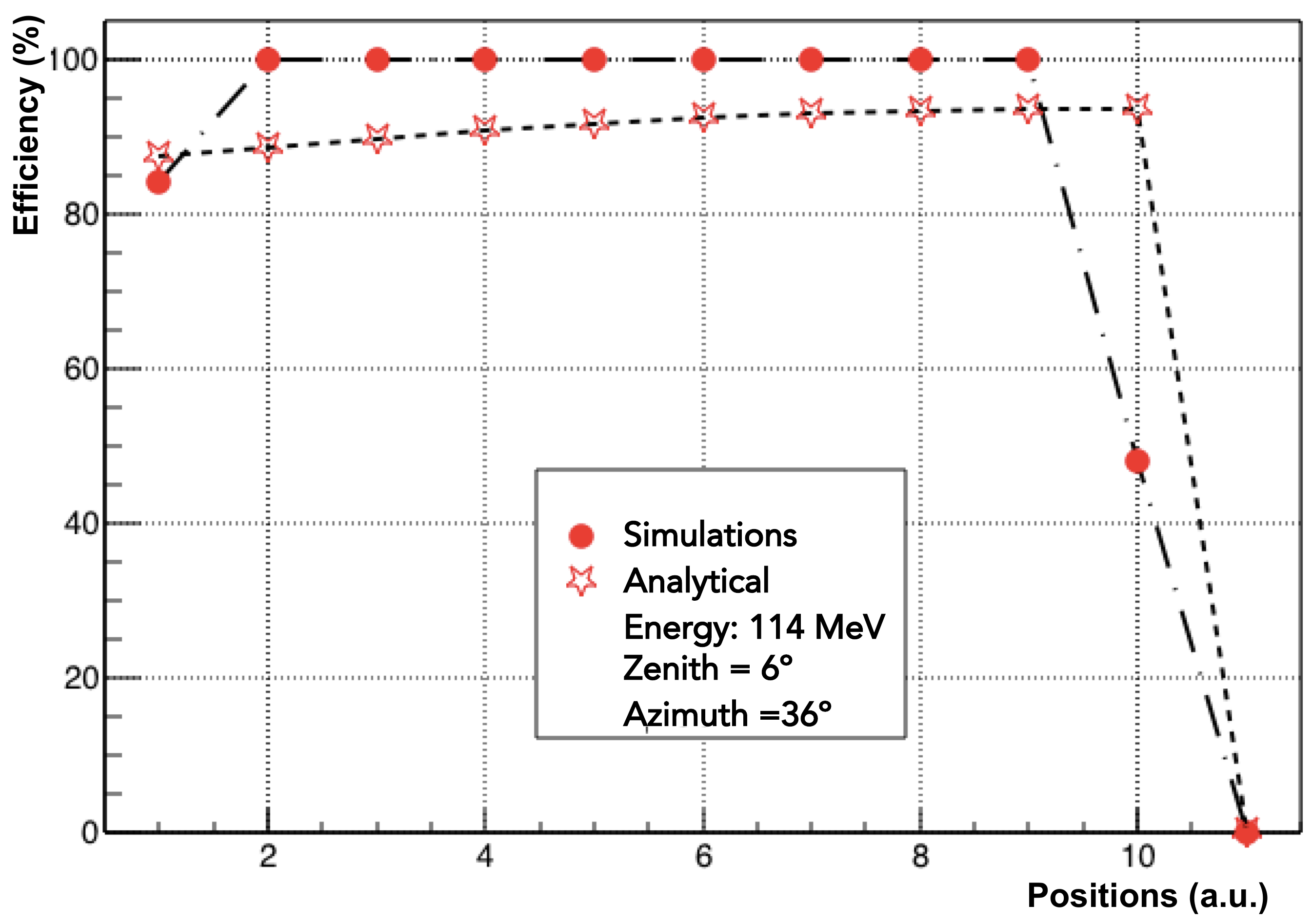}
\begin{center}
\vspace{-0.5 cm}
\caption{\label{BatataResults}Comparison of simulations with expected results from analytical calculations for certain azimuth and zenith angles as example \cite{Itzel}.}
\end{center}
\end{minipage} 
\vspace{-0.9 cm}
\end{figure}

\noindent \textbf{Inclined showers and Muon Component} For estimating the energy of inclined showers, parameter $N_{19}$, representing the shower size is defined. It is based on the muon density model prediction to fit the signals recorded by the SD with respect to a reference profile for the muon density at ground of proton showers of $10^{19}$ eV simulated with several hadronic models. In order to test the effectiveness of $N_{19}$ as estimator of the total number of muons relative to that in the reference distribution, a simulated $N_{19}$ is compared to the true ratio $R^{MC}_{\mu} = N^{true} _{\mu} /N{\mu,19}$, where $N^{true} _{\mu}$ is the true number of muons at ground. The mean value of the difference between $N^{MC} _{19}$ and $R^{MC} _{\mu}$ is shown in Figure~\ref{fig:MuonInclined}.  There is a bias of less than $ 5\%$ for showers with $R^{MC} _{\mu}  > 0.6$, value above which the SD array is over $ 95\%$ efficient. From this result, we conclude that $N_{19}$ provides a direct measurement of the relative number of muons with respect to the reference distribution \cite{MiInclinada}.
\vspace{-0.5 cm}
\subsection{Low Energy muon studies}
\label{Lowmuon}Burried Array Telescope at Auger ({\bf BATATA}) is a muon telescope built to analyze the muon component of EAS and to quantify the electromagnetic contamination as a function of the depth, for energies above $6 \cdot 10^{15}$  eV.  BATATA is conformed by plastic scintillators bars meant to validate the depth of the design of AMIGA counters. BATATA  is also an independent muon detector. A simulation code has been developed to reproduce the data of the detector for comparisons with measurements. For testing the code, results obtained with it are compared with the expected results from analytical calculations for 9 positions on the detector (positions 1 to 9 in  Figure~\ref{BatataResults}). This study allowed to characterize the use of the simulation code (for particle interactions in the EAS at lower energies than the primary one) for getting the most reliable results. The maximum difference is $\sim 10\%$ \cite{Itzel}.  
\vspace{-0.5 cm}
\subsection{Exotic Particles UHECRONS}
\label{Exotic}
 Proposed exotic particle called UHECRON \cite{IvoneAl} is studied through simulations and compared with measurements of the Pierre Auger Observatory. UHECRON is a strongly interacting, long lived, massive and neutral particle, which do not has to interact with the Cosmic Background Radiation being able to reach the earth. Hybrid events are selected and compared with simulated events, those are mixtures of proton-UHECRON with 0.5, 2.5, 5 and 10 $\%$ of UHECRON content.  Energies considered are from $10^{18}$ up to $10^{20}$ eV. The mixture with around 0.5 $\%$ of UHECRONs is selected as the one that best represents the experimental data.  A Bayesian approach in order to determine the upper limit for the UHECRON
composition fraction that may be present in the experimental data collected by the Observatory for energies above $10^{19}$ eV is performed. The upper limit for the UHECRON composition fraction determined by this
method is 4.4$\%$ at 95$\%$ C.L. \cite{Francisco}. For more details on this kind of studies see \cite{Schuster,Francisco}.
\vspace{-0.5 cm}
\ack{}
Many thanks to CONACyT Mexico  for financial support through projects CB 243290 and CB 132197, and L'OREAL Mexico, CONACyT, UNESCO, CONALMEX and AMC for the \textit{Fellowship for Women in Science 2014}.
I want to thank also to all collaborators, who contributed to the work described in this document.

\end{document}